\documentclass[sigconf]{acmart}

\usepackage{booktabs} % For formal tables
\usepackage{graphicx}
\setcopyright{none}
\setcopyright{none}
\acmDOI{}
\acmISBN{}
\acmPrice{}
\acmConference{}{}{}
\acmBooktitle{}

\AtBeginDocument{%
  \makeatletter
  \gdef\@acmConference@info{}%
  \gdef\@copyrightpermission{}%
  \renewcommand\footnotetextcopyrightpermission[1]{}%
  \gdef\@formatdoi#1{}%
  \pagestyle{plain}
  \makeatother
}
\begin{document}
\title{iblock: Accurate and Scalable Bitcoin Simulations with OMNeT++}
%\titlenote{Produces the permission block, and
 % copyright information}
%\subtitle{Extended Abstract}
%\subtitlenote{The full version of the author's guide is available as
 % \texttt{acmart.pdf} document}
  
%\renewcommand{\shorttitle}{SIG Proceedings Paper in LaTeX Format}

\author{Niccol\`o Scatena}
\affiliation{%
 \institution{Department of Information Engineering \\
 University of Pisa}
 \streetaddress{Via G. Caruso 16, 56122}
 \city{Pisa} 
 \country{Italy}
 \postcode{56122}  
}
\email{n.scatena1@studenti.unipi.it}
\author{Pericle Perazzo}
\affiliation{%
 \institution{Department of Information Engineering \\
 University of Pisa}
 \streetaddress{Via G. Caruso 16, 56122}
 \city{Pisa} 
 \country{Italy}
 \postcode{56122}  
}
\email{pericle.perazzo@unipi.it}
\author{Giovanni Nardini}
\affiliation{%
 \institution{Department of Information Engineering \\
 University of Pisa}
 \streetaddress{Via G. Caruso 16, 56122}
 \city{Pisa} 
 \country{Italy}
 \postcode{56122}  
}
\email{giovanni.nardini@unipi.it}
\renewcommand{\shortauthors}{N. Scatena et al.}

%\author{Anonymous Authors}

\begin{abstract}
This paper proposes iblock, a comprehensive C++ library for Bitcoin simulation, designed for OMNeT++.
iblock offers superior efficiency and scalability with respect to state-of-the-art simulators, which are typically written in high-level languages. 
Moreover, the possible integration with other OMNeT++ libraries allows highly detailed simulations. 
We measure iblock's performance against a state-of-the-art blockchain simulator, proving that it is more efficient at the same level of simulation detail. We also validate iblock by using it to simulate different scenarios such as the normal Bitcoin operation and the selfish mine attack, showing that simulation results are coherent with theoretical expectations.
\end{abstract}

%   <concept>
%     <concept_id>10002978.10003006.10003013</concept_id>
%     <concept_desc>Security and privacy~Distributed systems security</concept_desc>
%     <concept_significance>500</concept_significance>
%   </concept>
%   <concept>
%     <concept_id>10002950.10003648.10003662</concept_id>
%     <concept_desc>Mathematics of computing~Simulations</concept_desc>
%     <concept_significance>300</concept_significance>
%   </concept>
%   <concept>
%     <concept_id>10002978.10003029.10011703</concept_id>
%     <concept_desc>Security and privacy~Economics of security and privacy</concept_desc>
%     <concept_significance>300</concept_significance>
%   </concept>
%   <concept>
%     <concept_id>10002944.10011123.10011130</concept_id>
%     <concept_desc>General and reference~Experimentation</concept_desc>
%     <concept_significance>100</concept_significance>
%   </concept>

% \ccsdesc[500]{Security and privacy~Distributed systems security}
% \ccsdesc[300]{Mathematics of computing~Simulations}
% \ccsdesc[100]{Security and privacy~Economics of security and privacy}
% \ccsdesc[100]{General and reference~Experimentation}


\maketitle

\section{Introduction}
\label{sec:introduction}

Bitcoin~\cite{nakamoto2008bitcoin,antonopoulos2023mastering}, introduced in 2008 by an anonymous entity known as Satoshi Nakamoto, is a decentralized digital currency that operates without a central authority or intermediary. 
It relies on a peer-to-peer network, cryptography, and proof-of-work consensus to ensure transactions security and to maintain the integrity of the blockchain, which is a public ledger recording all transactions. 
Simulating the Bitcoin system is crucial for understanding its dynamics, scalability, and security. Through simulation, researchers can analyze the impact of various factors such as transaction volume, network latency, and mining difficulty on the overall performance and stability of the system. 
This helps in identifying potential vulnerabilities, optimizing protocols, and ensuring the robustness of Bitcoin against malicious attacks and systemic risks. 
Despite significant advancements, current state-of-the-art Bitcoin simulators~\cite{paulavicius2021systematic,androulaki2013evaluating,miller2015shadowbitcoin,gervais2016security,alharby2019blocksim,faria2019blocksim,basile2022segwit,stoykov2017vibes,aoki2019simblock} still face notable inefficiencies. 
These simulators often fail in accurately modeling the complex and dynamic nature of the Bitcoin system, particularly with many nodes and under high transaction volumes. 
Many simulators lack the ability to replicate aspects of real-world scenarios, such as variable transaction fees, transaction sizes, block sizes, and block intervals, which are important parameters for assessing the efficiency and resilience of the network. 
Additionally, due to the high-level languages in which Bitcoin simulators are written (Python, Java, Scala), the computational resources required for detailed simulations can be prohibitively high, thus limiting their scalability. 
These inefficiencies limit the possibility of conducting comprehensive analyses of the Bitcoin system.

This paper proposes iblock, a comprehensive C++ library and simulation model for the Bitcoin system, which enables simulations across various layers, i.e. network, consensus, incentives, data. 
Written in C++ and designed for OMNeT++~\cite{varga2001omnet}, iblock offers superior efficiency and scalability with respect to state-of-the-art simulators written in high-level languages. 
iblock provides in-depth representation of Bitcoin's data structures and algorithms, which facilitates experiments across consensus, incentive mechanisms and data management. 
Moreover, since OMNeT++ is a general-purpose simulator, the possible integration with other OMNeT++ libraries allows highly detailed simulations. 
iblock's modular and extensible architecture allows for the seamless addition of new features and behaviors, providing a valuable framework for studying their effects on the Bitcoin system. 
We measure iblock's performance against BlockSim~\cite{alharby2019blocksim}, which is a state-of-the-art blockchain simulator, proving that iblock is more efficient at the same level of simulation detail. 
We also validate iblock's accuracy by comparing simulation results with theoretical models and related studies, and showcase iblock's capability to simulate different scenarios such as the selfish mine attack~\cite{eyal2013majority}.
We make the iblock's code publicly available through github\footnote{https://github.com/SpeedJack/iblock}

The rest of the paper is organized as follows. 
Section~\ref{sec:related_work} presents related work and compares with it. 
Section~\ref{sec:background} introduces some necessary background about Bitcoin and OMNeT++. 
Section~\ref{sec:iblock} describes iblock and its internal structure. 
Section~\ref{sec:proofs_of_concept} validates iblock with some proof-of-concept simulations, namely the Bitcoin normal operation and a selfish mine attack. 
Section~\ref{sec:performance} experimentally analyzes its scalability and compares its performance against the BlockSim simulator. 
Finally, Section~\ref{sec:conclusions} concludes the paper.

\section{Related Work}
\label{sec:related_work}

Several blockchain simulators have been proposed in the literature~\cite{paulavicius2021systematic}. 
The first Bitcoin simulator, to the best of our knowledge, has been the `Bitcoin privacy simulator' by Androulaki et al.~\cite{androulaki2013evaluating}, which however simulated only few aspects of the Bitcoin ecosystem, namely those related to addresses. 
Shadow-Bitcoin~\cite{miller2015shadowbitcoin} is not strictly speaking a simulator but rather a process-level virtualization system capable of running directly the reference Bitcoin implementation over a virtual OS and a virtual network. 
Consequently, it achieves extreme realism at the cost of inefficiency and usage complexity. 
At the time of its publication, Shadow-Bitcoin did not support the SegWit upgrade.

Bitcoin-Simulator~\cite{gervais2016security}, built upon NS-3, is maybe the most similar proposal to ours, and it reaches comparable efficiency. 
It simulates variable transaction and block size, variable block interval, and peer-to-peer communication. 
However, it still fails to simulate many aspects like transaction fees and difficulty adjustments. 
It does not support the SegWit upgrade, which was not yet deployed on the Bitcoin network at the time.

BlockSim by Alharby et al.~\cite{alharby2019blocksim} and the homonymous BlockSim by Faria et al.~\cite{faria2019blocksim} are both realized in the Python language. 
The latter is built upon the SimPy library, while the former is built from scratch. 
None of them originally supported SegWit, although the BlockSim by Alharby et al. has been extended in that sense~\cite{basile2022segwit}. 
Since they are written with a high-level language, they do not achieve the same efficiency of C++-based simulators as the one proposed by the present paper. 
Both BlockSims fail to simulate the variable transaction and block sizes, the dynamic difficulty adjustments, and the UTXOs. 
In Section~\ref{sec:performance} we quantitatively compare the efficiency of iBlock with BlockSim by Alharby et al., showing that we achieve both better performance and higher realism.
BlockPerf~\cite{polge2021blockperf} by Polge et al. extends the Faria et al.'s BlockSim by adding a real network level implemented in C/C++, but it offers performances even worse than the baseline BlockSim.
VIBES~\cite{stoykov2017vibes} and SimBlock~\cite{aoki2019simblock} simulators are among the first to support the SegWit upgrade, but they still miss some important aspects like variable transaction and block sizes and UTXOs. 
Moreover, they still use inefficient high-level language, namely Scala and Java respectively.
BSELA by Cui and Hu~\cite{cui2024bsela} simulates Bitcoin with an event-based architecture, which adopts a mix between queues and min-heaps as data structures to maintain intra- and inter-block events. 
This may improve efficiency if applied to OMNeT++, which has its own built-in mechanism to maintain simulated events. 
Such an optimization is orthogonal to ours, and we leave this integration as a possible future work.

Table~\ref{tbl:features} summarizes the comparison of our proposed simulator with relevant others in the literature.

\begin{table*}
    \caption{Feature comparison with relevant other Bitcoin simulators}
    \label{tbl:features}
    \centering
    \begin{tabular}{c|ccccccccc}
% $\times$ \checkmark
     & Var. tx fee & Var. tx size & UTXOs & Var. block size & Var. block interval & Diff. adjust. & Scalability \\
     \hline
    `Bitcoin privacy simulator' \cite{androulaki2013evaluating} & $\times$ & $\times$ & $\times$ & $\times$ & $\times$ & $\times$ & \checkmark \\
    %Shadow-Bitcoin \cite{miller2015shadowbitcoin} & \checkmark & \checkmark & \checkmark & \checkmark & \checkmark & \checkmark & \checkmark & $\times$ \\
    Bitcoin-Simulator \cite{gervais2016security} & $\times$ & $\times$ & $\times$ & \checkmark & \checkmark & $\times$ & \checkmark \\
    BlockSim (Alharby) \cite{alharby2019blocksim} & \checkmark & \checkmark & $\times$ & $\times$ & $\times$ & $\times$ & $\times$ \\
    BlockSim (Faria) \cite{faria2019blocksim} & $\times$ & $\times$ & $\times$ & $\times$ & $\times$ & $\times$ & $\times$ \\
    VIBES \cite{stoykov2017vibes} & $\times$ & $\times$ & $\times$ & $\times$ & $\times$ & $\times$ & $\times$ \\
    SimBlock \cite{aoki2019simblock} & $\times$ & $\times$ & $\times$ & $\times$ & $\times$ & $\times$ & $\times$ \\
    iblock & \checkmark & \checkmark & \checkmark & \checkmark & \checkmark & \checkmark & \checkmark \\
\end{tabular}
\end{table*}

\section{Background}
\label{sec:background}

\subsection{Bitcoin}
Bitcoin~\cite{nakamoto2008bitcoin,antonopoulos2023mastering} is a decentralized digital currency that allows for peer-to-peer transactions without a bank or other intermediaries. 
It was introduced in 2008 by an anonymous person or group of people using the pseudonym Satoshi Nakamoto. 
Bitcoin is based on a public ledger called the `\emph{blockchain}', which records all transactions that have occurred on the network. 
The blockchain is a chain of blocks, where each block contains some metadata plus a list of transactions. 
Blocks are chained in chronological order, with each block containing a hash to the previous block. 
Therefore, changing a transaction in any block would require changing all subsequent blocks. 
Since computing each block requires solving a difficult cryptographic puzzle, this makes the blockchain tamper resistant. 
This technique is called `Proof-of-Work' (\emph{PoW}) consensus.
A peer-to-peer network of nodes communicate to each other new transactions and new blocks, and some of them (\emph{miners}) also compete in computing new blocks by solving the cryptographic puzzle. 
Each miner that succeeds in computing a new block is rewarded with a quantity of digital coins, both created from scratch (\emph{block subsidy}) and paid as \emph{fees} by the issuers of the transactions included in the block. 
The difficulty of the cryptographic puzzle is automatically tailored every 2016 blocks (\emph{difficulty adjustment}) in such a way to maintain the average time between successive created blocks as close as possible to 10 minutes.

Transactions are data structures used to transfer coins between users. 
They are signed by the payer, so that they prove the willingness of the payer to pay a specific amount to the recipient. 
Both payer and recipient use \emph{wallets}, which are pieces of software that generate and store pairs of public and private keys used for signing transactions. 
Each transaction has multiple inputs, where the coins are taken from, and multiple outputs, where the coins are sent to. 
Each input references an output of a previous transaction and it spends it, while each output specifies the amount of money to be sent and one public key of the receiver. 
Until spent, the output of a transaction is called `Unspent Transaction Output' (\emph{UTXO}). 
Transactions are broadcast to the peer-to-peer network, and each node adds them to a data structure called `\emph{mempool}'. 
The mempool contains only \emph{unconfirmed} transactions, meaning that they have not yet been included in any block. 
Each miner selects transactions from its mempool, usually by choosing those with the highest ratio between fee and transaction size and includes them in the block that is trying to solve the cryptographic puzzle of. 
When a new block is created and broadcast by some miner, all the nodes add such a  block to their locally maintained blockchain and remove the relative transactions from their mempool.

\subsection{OMNeT++}
OMNeT++ is an open-source, modular simulation framework written in C++. Although it was primarily designed to simulate networks, it is general enough to support the modeling of any system that can be represented according to the discrete-event dynamic simulation (DEDS) paradigm. Indeed, OMNeT++ comes with a simulation engine that provides all things useful to create simulation models, such as event queues, event schedulers, random number generation, etc., hence allows the user to focus on developing the logic of the desired simulation model only. Nonetheless, OMNeT++ facilitates the implementation of object-oriented simulation models by encouraging developers to follow basic implementation principles.

\begin{figure}[t]
  \centering
  % [trim=left bottom right top, clip]
  \includegraphics[width=0.9\linewidth]{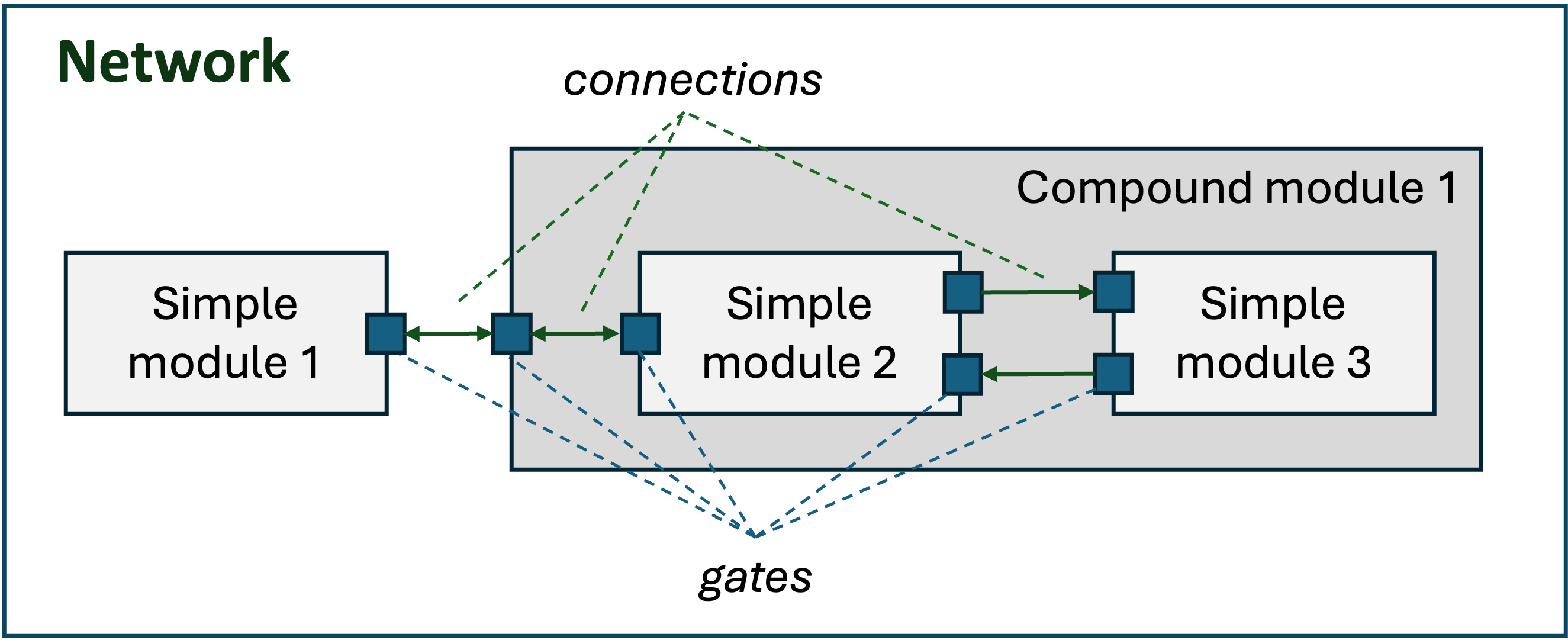}
  \caption{OMNeT++ module hierarchy.}
  \label{fig:omnetpp}
\end{figure}

The building blocks of an OMNeT++-based simulation model are shown in Fig.~\ref{fig:omnetpp}: \textit{modules} - which can be either simple or compound - communicate via \textit{messages}, which are usually sent and received through a \textit{connection} - characterized by a latency and, where applicable, a bit rate and loss probability - linking their \textit{gates}, which act as interfaces. The reception of messages at a module are the events that let the simulation state and time evolve. Modules can be grouped together to form a \textit{network}, that is the system to be simulated. Model description, implementation and parametrization are kept separate, and specified using different languages. Modules are described using the Network Description (NED) language, which allows the user to define the structure of a module in terms of parameters, gates, submodules (if any) and internal connections. A C++ class is used to specify the \textit{behavior} of a module, i.e., how it reacts to the reception of messages. Once a network has been composed, initialization (INI) files are used to assign actual values to the modules' parameters. This is important because different simulation scenarios can be configured by simply changing parameter values in the INI file, without modifying either the module's behavior (i.e., the C++ class) or its structure (i.e., the NED file). The above separation between model definitions and parameters allows the user to automate the creation of simulation studies: the latter are generated from INI files, by automatically computing the Cartesian product of all the factors (i.e., parameters which are assigned a set of values) and generating one simulation instance for each combination. Independent replicas of the same instance can also be run automatically using different seeds, which will be fed to the random number generators. 

\begin{figure}[t]
  \centering
  % [trim=left bottom right top, clip]
  \includegraphics[trim=0 3.97cm 9cm 0, clip, width=\linewidth]{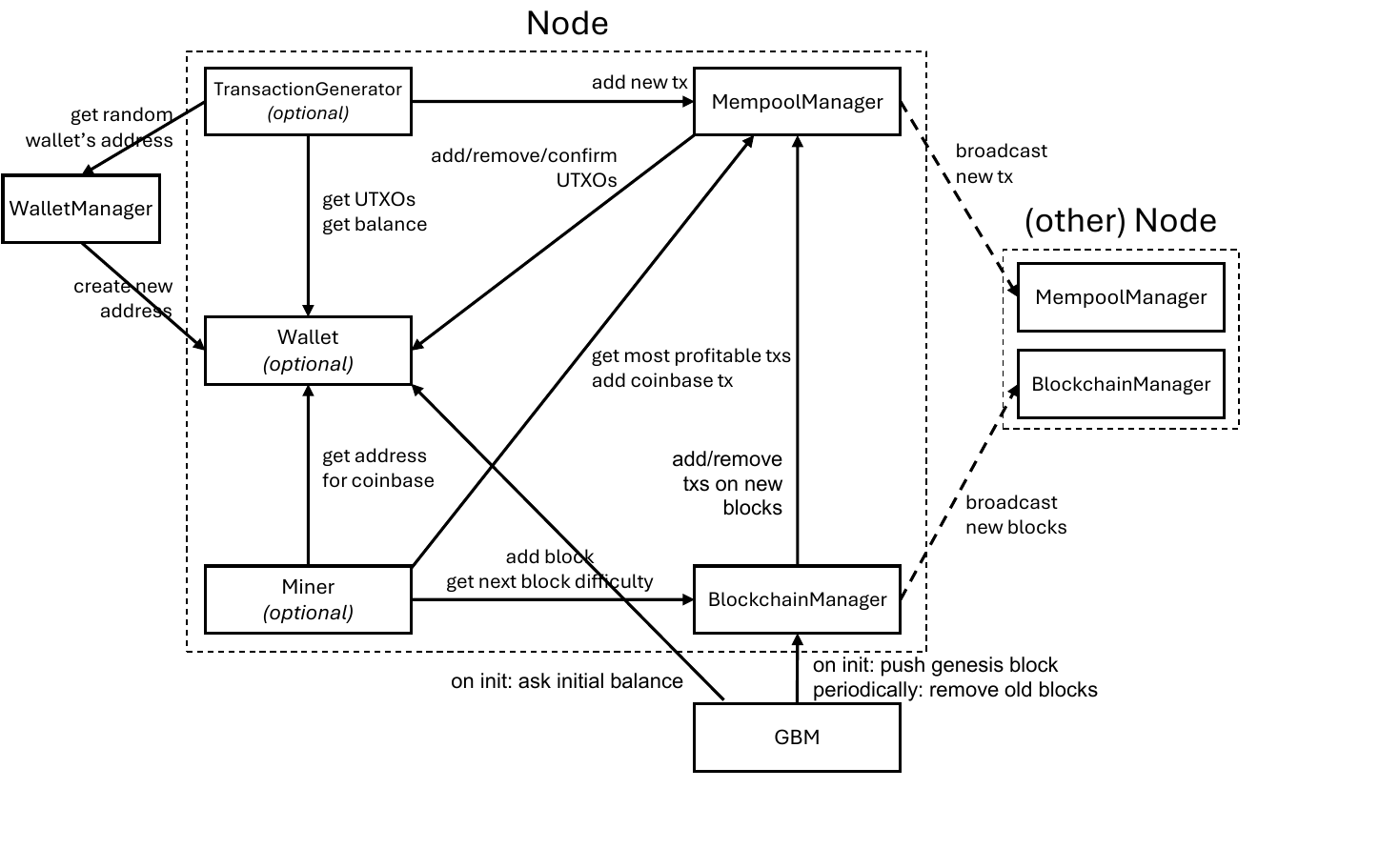}
  \caption{iblock node architecture. Solid arrows indicate DMCs, while dashed ones indicate direct messages.}
  \label{fig:applications}
\end{figure}

Finally, OMNeT++ comes with a wide variety of simulation libraries developed by the community. Notable examples are: the INET library \cite{inet}, which provides models for TCP/IP protocols and devices (e.g., routers and switches); Simu5G \cite{simu5g}, which models 4G and 5G cellular infrastructure; Artery \cite{artery}, which models vehicular networks' communications. Existing models like the ones above can be composed to simulate arbitrarily complex simulation scenarios.
\section{iblock}
\label{sec:iblock}

iblock is a C++ library and simulation model for the Bitcoin system, which enables simulations across various layers, i.e. network, consensus, incentives, data. 
It is developed in C++23 on top of the OMNeT++ discrete-event simulation framework and is compatible with OMNeT++ version 6.x. 
An iblock network consists of multiple nodes that communicate through direct messages. 
Nodes in iblock can vary by type and are configured with different parameters, utilizing the flexibility of OMNeT++'s configuration capabilities. 
Miners contain a Miner application which simulates the actions of mining blocks and earning rewards, while user nodes feature a TransactionGenerator application which simulates the creation and submission of transactions to the network. 
The network also includes special-purpose modules, namely the Global Blockchain Manager (GBM) module, which creates the genesis block at startup and manages memory by removing unneeded blocks, the WalletManager module, which acts as a directory for other nodes to retrieve wallet addresses, and the NodeManager module, which serves as a directory for all nodes in the network, allowing nodes to request references for message sending. 
Nodes in iblock are compound modules composed of various applications, each of which is a simple module that handles a specific task. 
All applications inherit from the AppBase class, which provides foundational features, making it easier to add new functionalities. 
Nodes in iblock are highly modular: users can tailor a node's capabilities by selecting different applications, allowing each node to exhibit unique functionalities. 
Since a Bitcoin node maintains its own state defined by its local view of the blockchain and its mempool, every iblock node includes two essential applications: the BlockchainManager and the MempoolManager. 
%Additional applications can be added to customize node behavior using NED language. 
%Each iblock node maintains its own perspective of the blockchain and mempool, just like nodes in a real Bitcoin network. 
This means one node's blockchain or mempool may differ from another's, as nodes may have different blocks or transactions at any point in time. 
Fig.~\ref{fig:applications} shows the software architecture of a node, and the interactions within applications within it and across different nodes.
The applications tagged with ``(optional)'' may be absent, depending on the node's capabilities. 
In particular, the Wallet application is needed only if the node maintains its own UTXOs, the TransactionGenerator only if the node can spend and receive bitcoins, and the Miner applications only if the node participates to the PoW consensus. 
The Miner and the TransactionGenerator applications both needs also the Wallet application, respectively to receive the mining rewards and to spend owned UTXOs. 
The GBM and the WalletManager are global modules.
Internal communication between applications and with global modules uses Direct Method Calls (DMCs), while node-to-node communication relies on direct messages. 
The figure illustrates key interactions between applications within a node. 
The BlockchainManager and the MempoolManager applications are the only two mandatory applications, which define a full node. 

\subsection{The BlockchainManager Application}
The BlockchainManager application maintains the local blockchain view for each node. 
This module is capable of simulating forks, chain reorganizations, and orphan blocks. 
When a node mines a new block, it broadcasts it to all other BlockchainManager instances in the network. 
To do this, it uses the NodeManager global module to retrieve references to all the other nodes.
BlockchainManager simulates a different block propagation time for each node, as the block is modeled as an OMNeT++ message received with different random delays by destination nodes.
This allows us to mimic the behavior of the gossip protocol while avoiding the burden of modeling the entire message exchange, which would otherwise increase the computational cost of the simulator and reduce its efficiency.
Upon receiving a new block, BlockchainManager extends the main branch if the block builds on it. 
This involves notifying MempoolManager to remove included transactions and updating Wallet to possibly adjust the confirmation count of UTXOs. 
BlockchainManager also discards possible duplicate blocks and handles possible forks, that is, if the new block extends a non-main branch, it checks whether this branch is now the longest one. 
If so, it performs a reorganization in which the main branch is reverted backward to the fork point, updating the MempoolManager and unconfirming UTXOs in Wallet. 
The new branch is then added as the main branch, block by block. 
Finally, at every epoch (i.e., 2016 blocks), BlockchainManager recalculates the mining difficulty based on the time needed to mine the epoch blocks. 
To save space in the simulator, each distinct block is represented by a unique data structure in the GBM module, which is common for all the BlockchainManagers. 
Each BlockchainManager maintains only the pointers to such data structures, or more precisely the points to the branches' head blocks.
At simulation start, the GBM adds the genesis block to each BlockchainManager. 
Moreover, the GBM periodically calls the cleanup method of each BlockchainManager to prune unused branches, thus freeing memory. 

\subsection{The MempoolManager Application}
The MempoolManager handles the mempool by maintaining an ordered list of transactions based on fee rate, allowing miners to prioritize transactions for maximum profitability. When a node creates a new transaction, it broadcasts the transaction to each MempoolManager instance in the network using direct messages. 
To do this, it uses the NodeManager global module to retrieve references to all the other nodes.
MempoolManager simulates a different transaction propagation time for each node, as the transaction is modeled as an OMNeT++ message received with different random delays by destination nodes.
Upon receiving a new transaction, MempoolManager adds it to its mempool and notifies Wallet if any new UTXO is available. 

\subsection{The Wallet Application}
The Wallet application manages the node's UTXOs. 
Each time a new block is added to the local blockchain or a transaction is added to the mempool, the Wallet updates its UTXO list accordingly. 
When a new transaction is added to the mempool, the MempoolManager checks whether any transaction outputs match the Wallet's address. 
If so, it adds UTXOs in the Wallet. When a block is added to the blockchain's main branch, Wallet increases the UTXO confirmation count, while when a UTXO is spent by a transaction in the mempool, it removes such a UTXO.

\subsection{The Miner Application}
The Miner application is responsible for generating new blocks. 
At the start of the simulation, a timer is scheduled to trigger mining operations. 
When the timer activates, the Miner produces a new block at the top of the current main blockchain branch and then instructs the BlockchainManager to add this block to the chain and distribute it across the network. 
As part of this process, the Miner also creates a coinbase transaction to claim the block reward, whose output is directed to a particular address specified by the Wallet application.

\subsection{The TransactionGenerator Application}
The TransactionGenerator application is responsible for creating transactions with fully configurable parameters, including amount, fee, number of outputs, generation interval, and more. Before generating a transaction, TransactionGenerator verifies with the Wallet application that sufficient balance is available. 
If funds are insufficient, TransactionGenerator waits until the balance meets the required amount. 
Once verified, it retrieves the list of UTXOs for the transaction from the Wallet and requests one or more randomly selected recipient addresses from the WalletManager global module. 
To handle the change, an additional transaction output is created using the node's Wallet address. Finally, the transaction is added to the mempool via the MempoolManager, which in turn broadcasts it to the other nodes.

\section{Validation and Proofs of Concept}
\label{sec:proofs_of_concept}

This section provides two examples of simulations that can be conducted using iblock. 
\subsection{Bitcoin Normal Operation}
First of all, we present a simulation of a Bitcoin network configured with miners and non-miner nodes, which create blocks, generate transactions, and propagate them throughout the network. 
This simulation serves as a proof of concept but also as a validation of the proposed model. 
The network model has the following characteristics:
\begin{itemize}
\item 50 non-miner nodes that create transactions at an average rate of one every 20 seconds;
\item 10 miners, each with varying hash rates (namely 0.1\%, 0.4\%, 0.5\%, 1\%, 3\%, 5\%, 10\%, 15\%, 25\%, and 40\%), also generating transactions at the same average rate;
\item One non-miner node generating transactions at a slower rate of one every 5 minutes;
\item One non-miner node generating transactions at a faster rate of one every second.
\end{itemize}
In total, the network consists of 62 nodes, producing slightly over 4 transactions per second globally. This corresponds to the mean rate of transactions added to the mempool before 2023 in the real Bitcoin network, which was approximately 4-5 transactions per second\footnote{https://www.blockchain.com/explorer/charts/transactions-per-second.}. The simulation duration is set to 3 days, with 30 independent runs performed using different random seeds.

\begin{figure}[t]
  \centering
  % [trim=left bottom right top, clip]
  \includegraphics[trim=1.5cm 1.5cm 1.5cm 1.5cm, clip, width=\linewidth]{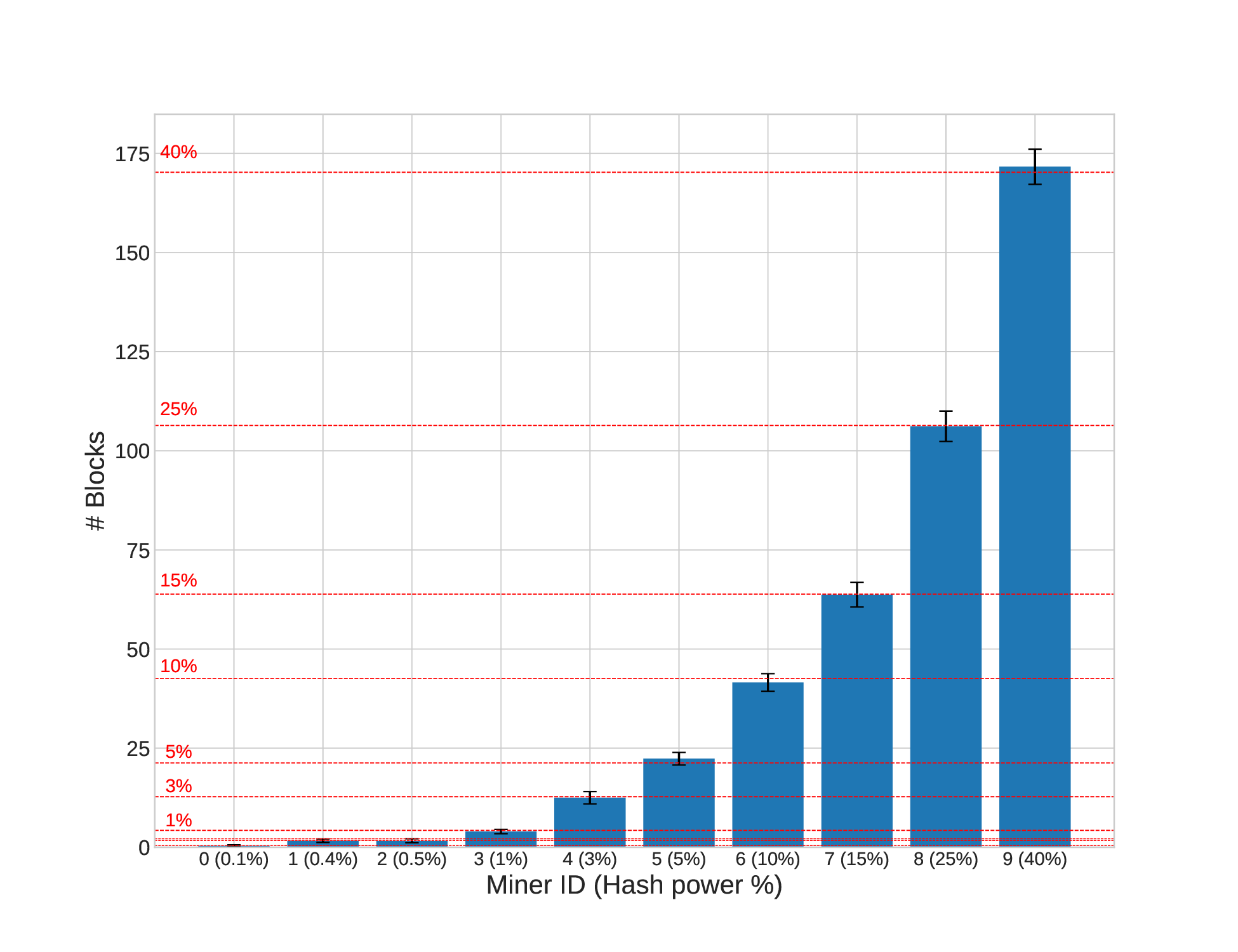}
  \caption{Number of blocks mined by each miner. Confidence intervals at 95\% level, with 30 samples. Red lines represent expected theoretical values.}
  \label{fig:honest_mined_blocks}
\end{figure}

Fig.~\ref{fig:honest_mined_blocks} illustrates the number of blocks mined by each miner, averaged over the 30 runs. 
The red horizontal lines represent the expected numbers of blocks per miner, calculated analytically. 
Notably, all 95\% confidence intervals include the relative expected numbers of blocks, confirming simulation consistency with theoretical values. 
The total number of blocks mined is 425.57, close to the expected 432 blocks based on the average of six blocks per hour that the Bitcoin protocol enforces. 
This indicates that iblock accurately models the PoW mining process within the Bitcoin system.

\begin{figure}[t]
  \centering
  % [trim=left bottom right top, clip]
  \includegraphics[trim=1.5cm 1.5cm 1.5cm 1.5cm, clip, width=\linewidth]{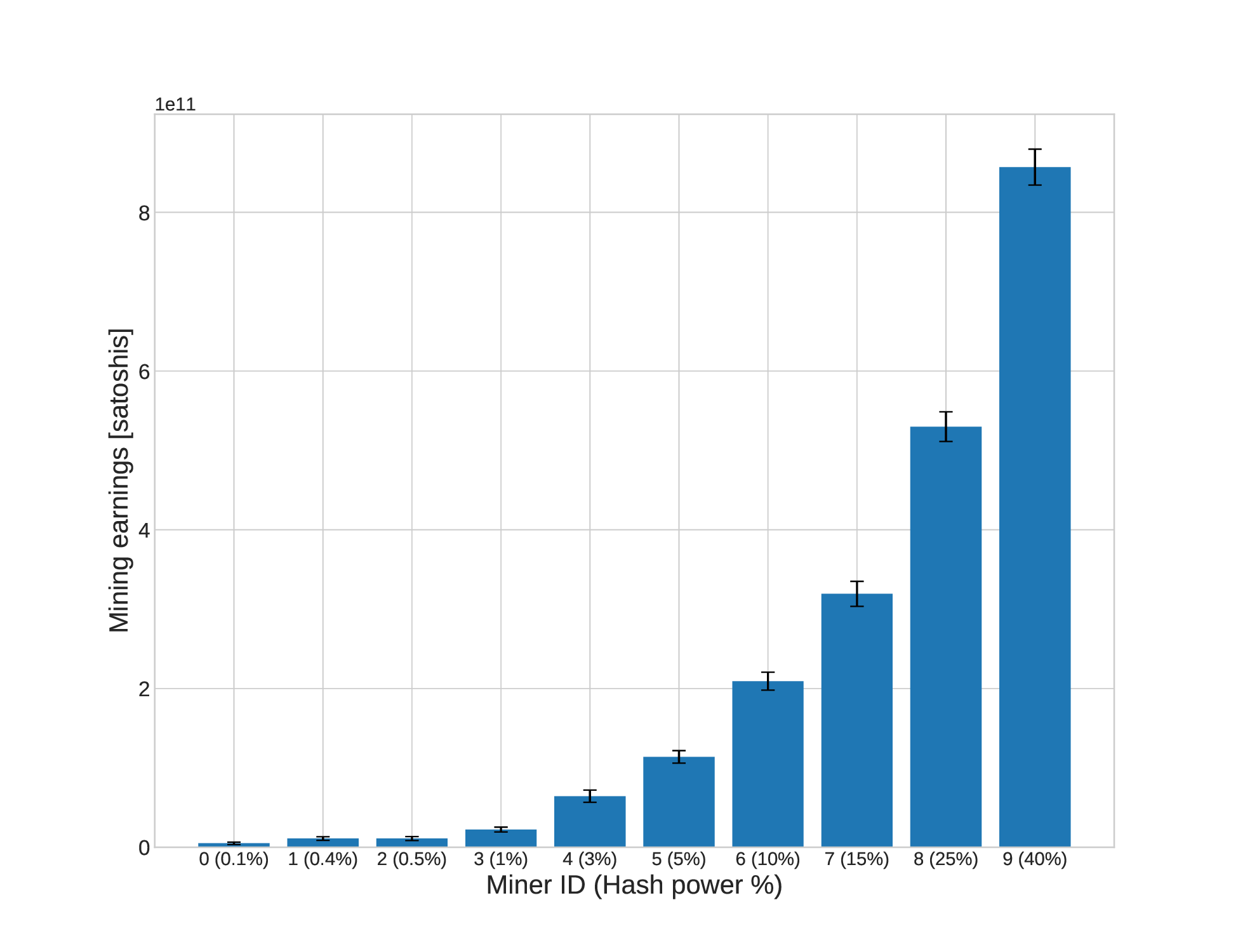}
  \caption{Mining rewards (in satoshis) per miner, including block subsidies and fees.}
  \label{fig:honest_rewards}
\end{figure}

Fig.~\ref{fig:honest_rewards} shows the mining rewards (in satoshis) per miner.
Of course, miners with higher hash rates mined more blocks and, consequently, received greater rewards. 
Therefore, these graphs resemble the distribution seen in Fig.~\ref{fig:honest_mined_blocks}.

\subsection{Selfish Mine}
Secondly, we present a simulation of a selfish mine attack~\cite{eyal2013majority} on the Bitcoin system. 
The attack consists of a malicious miner (the selfish miner) that mines blocks secretly, that is without broadcasting them to the network. 
The selfish miner will let other miners mine on the public blockchain and, in the meantime, it will mine on its own private chain. 
If the selfish miner manages to produce secretly two blocks in a row, from that moment the other miners are wasting their hash power. 
Indeed, even if they manage to produce a block, the selfish miner will suddenly reveal its two blocks, thus excluding the block produced by the other miners from the longest chain. 
If its hashing power is enough, the selfish miner is able to earn more rewards than what its hashing power should permit. 
Also, it could mount double-spending attacks even with less than 50\% hash power share.

For this analysis, we used two configurations. 
The first one includes a selfish miner, and it has the following characteristics:
\begin{itemize}
\item 45 non-miner nodes generating a transaction every 15 seconds in average;
\item 4 honest miners each contributing 20 TH/s to the total network's hash rate and creating transactions at the same rate;
\item One selfish miner node, with varying hash rate, also generating transactions at the same rate.
\end{itemize}
In total, the network consists of 50 nodes, producing about 3.33 transactions per second globally. 
The selfish miner is configured with a variable hash rate, respectively 10\%, 20\%, 30\%, and 40\% of the total network hash rate. 
It is simulated as a normal node, but replacing the standard BlockchainManager application with a special-purpose SelfishBCManager one, which implements the attack.
This demonstrates the flexibility of the iblock simulator.
Each of the six configurations has been run for 2 days of simulated time, and averaged over 20 independent repetitions.
The second configuration is for comparison, and it is identical to the first one, except that the selfish miner is replaced with an honest miner with the same varying hash rate. 
Also this simulation has been run for 2 days of simulated time, and averaged over 20 independent repetitions.

Table~\ref{tbl:successful_attacks} shows the number of successful attacks, that is the number of blocks mined by the selfish miner that have been included in the final blockchain, at different hash rate levels. 
\begin{table}
    \caption{Mean number of successful selfish mine attacks}
    \label{tbl:successful_attacks}
    \centering
    \begin{tabular}{c|c}
    Selfish miner's relative hash rate & Mean number of successful attacks \\
     \hline
    10\% (8.888 TH/s) & 3.35 \\
    20\% (20 TH/s) & 9.80 \\
    30\% (34.286 TH/s) & 14.55 \\
    40\% (53.333 TH/s) & 15.30 \\
\end{tabular}
\end{table}
Naturally, a higher hash rate results in a higher number of successful attacks. We observe also that this relationship is over-linear, confirming that the malicious strategy is convenient over the honest one.

\begin{figure}[t]
  \centering
  % [trim=left bottom right top, clip]
  \includegraphics[trim=1.5cm 1.5cm 1.5cm 1.5cm, clip, width=\linewidth]{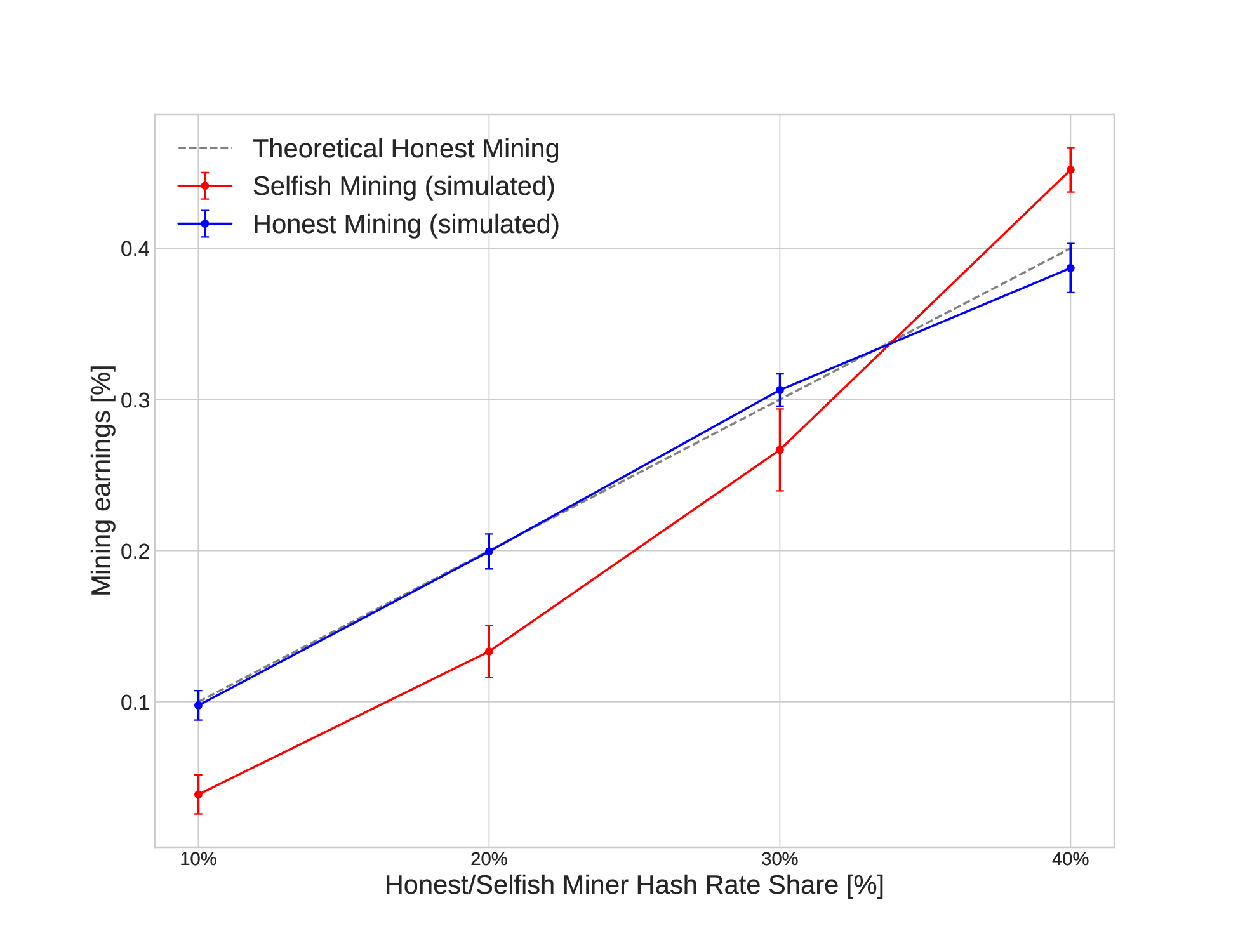}
  \caption{Number of blocks mined by each miner. Confidence intervals at 95\% level, with 30 samples. Red lines represent expected theoretical values.}
  \label{fig:selfish_rewards}
\end{figure}

Fig.~\ref{fig:selfish_rewards} shows the rewards earned by the miner with varying hash rate, operating either honestly or selfishly, as a percentage of total network rewards. 
The gray dashed line shows the theoretical expected reward for honest mining with each hash rate level. 
This is, basically, the identity function because the expected rewards are proportional to the hash rate. 
The blue and the red lines represent the average reward of respectively the honest and the selfish miners, with 95\% confidence intervals, as obtained from the simulation. 
The results indicate that the honest miner consistently receives expected rewards, as the blue line's confidence intervals always overlap with the gray dashed line, providing further validation of the accuracy of the iblock implementation. 
For all hash rates below 33\%, the red line remains below the gray dashed line (and below the blue line too), reflecting lower earnings for selfish mining. However, once this threshold is surpassed, the selfish miner's earnings exceed those of the honest miner. 
This result aligns with findings reported by Eyal and Sirer~\cite{eyal2013majority} with a propagation factor $\gamma=0$, which means that honest miners always avoid supporting the selfish miner's chain when it lacks a lead. 
This provides further validation of the correctness of the iblock implementation.

\section{Performance Evaluation}
\label{sec:performance}

This section presents two contributions. 
First, it provides insights on the scalability of iblock by evaluating how the execution time and memory usage vary when varying simulation parameters. 
Second, it compares the performance of iblock against the BlockSim simulator~\cite{alharby2019blocksim}.  
Simulations were run on a general-purpose computer equipped with an Intel Core i7-8705G CPU at 4.1 GHz, 16 GB of RAM and Linux operating system.

\subsection{Scalability Analysis}

In this section, we evaluate the scalability of iblock in terms of execution time and memory usage with varying network size and transaction rates. We first make the two factors vary one by one in order to assess their individual impact, then we make them vary together to assess their combined effect. The following results were obtained from simulations having a duration of three (simulated) days.

\begin{figure}[t]
  \centering
  % [trim=left bottom right top, clip]
  \includegraphics[trim=1.5cm 1.5cm 1.5cm 1.5cm, width=0.9\linewidth]{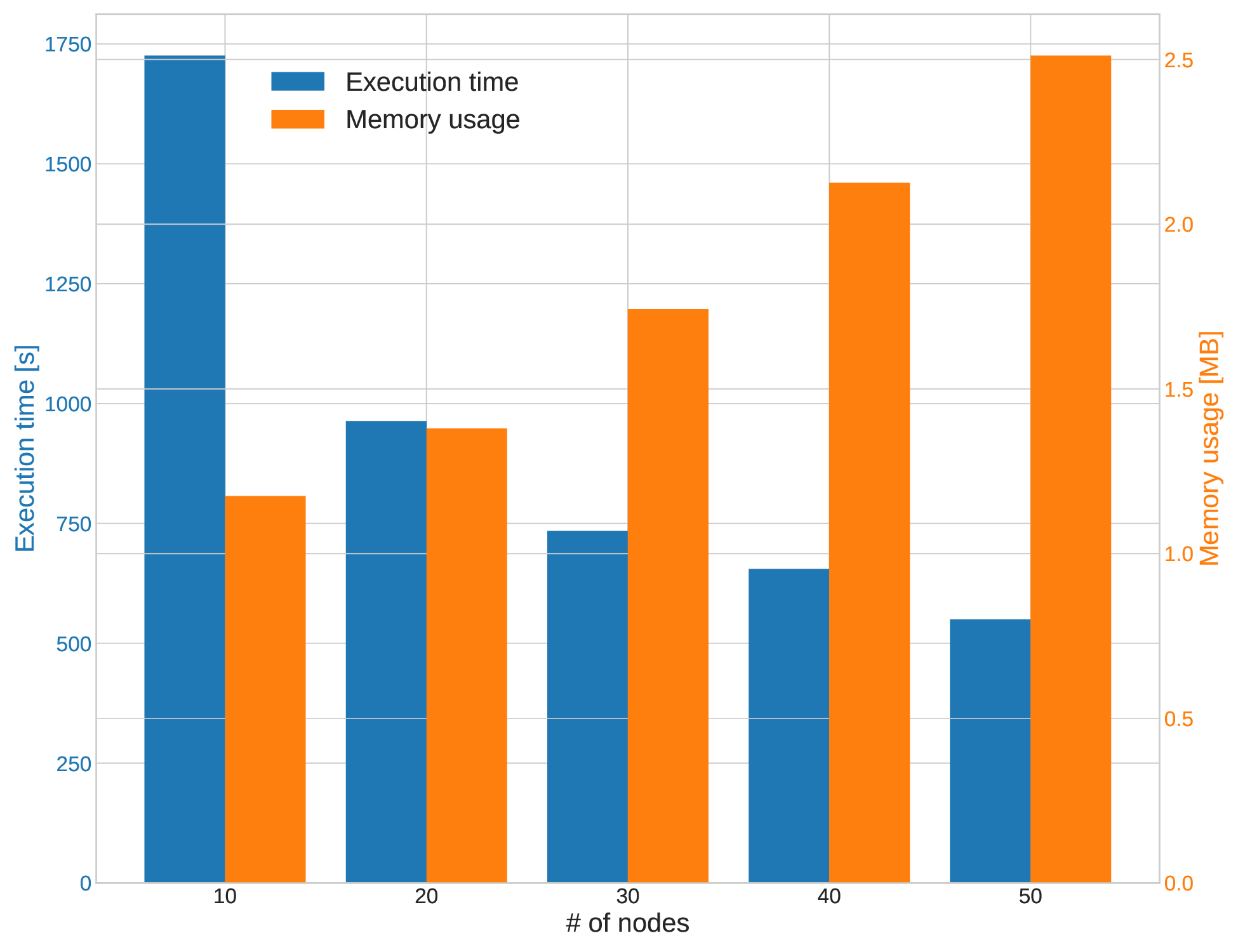}
  \caption{Execution time (primary y-axis, in blue) and memory usage (secondary y-axis, in orange) with varying number of nodes.}
  \label{fig:scalability_nodes}
\end{figure}

We first analyze the considered metrics when the number of nodes in the network increases - ranging from 10 to 50 nodes -, while keeping the global transaction generation rate constant at three transactions per second. Despite this scenario does not represent a realistic one (where, instead, the global transaction rate increases with the network size - this case will be analyzed later), it allows us to evaluate the performance of the simulator when varying the network nodes parameter in isolation. We assume that the number of miners is 20\% of the total number of nodes. 
Fig.~\ref{fig:scalability_nodes} shows that the execution time (reported in seconds on the primary y-axis) decreases as more nodes are considered. 
Although this may seem counterintuitive, it comes from the fact that the same number of UTXOs is distributed across more nodes, hence each node has fewer transactions to scan when searching for available funds, reducing its execution time.
As far as memory usage is concerned (shown in Megabytes on the secondary y-axis of Fig.~\ref{fig:scalability_nodes}), it increases with the number of nodes as expected. 
This is because each one has to maintain its own version of the mempool, hence consuming memory. 

\begin{figure}[t]
  \centering
  % [trim=left bottom right top, clip]
  \includegraphics[trim=1.5cm 1.5cm 1.5cm 1.5cm, width=0.9\linewidth]{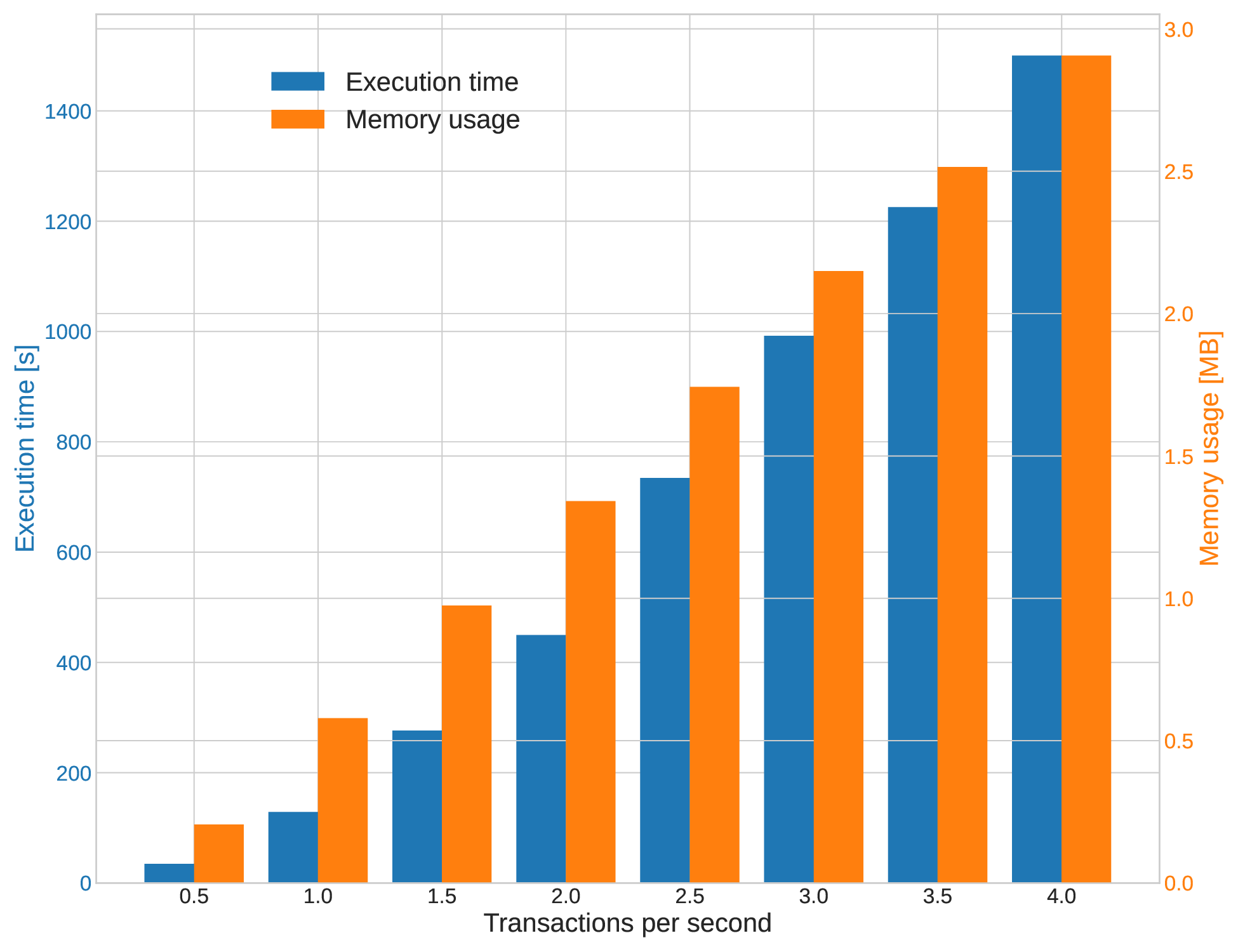}
  \caption{Execution time (primary y-axis, in blue) and memory usage (secondary y-axis, in orange) with varying transaction rate.}
  \label{fig:scalability_txs}
\end{figure}

We now show how the execution time and memory usage vary as the transaction rate increases while keeping the number of nodes fixed. 
In particular, we consider a network composed of 30 nodes and 6 miners. 
Simulation results are reported in Fig.~\ref{fig:scalability_txs}. 
In this case, the execution time increases with the transaction rate. 
On one hand, this is due to the higher number of transactions that the network must process. 
However, fund verification becomes more time-consuming as the UTXO set of each node becomes larger. 
Likewise, memory usage increases because each node must handle larger mempool and UTXO set.

\begin{figure}[t]
  \centering
  % [trim=left bottom right top, clip]
  \includegraphics[trim=1.5cm 1.5cm 1.5cm 1.5cm, width=0.9\linewidth]{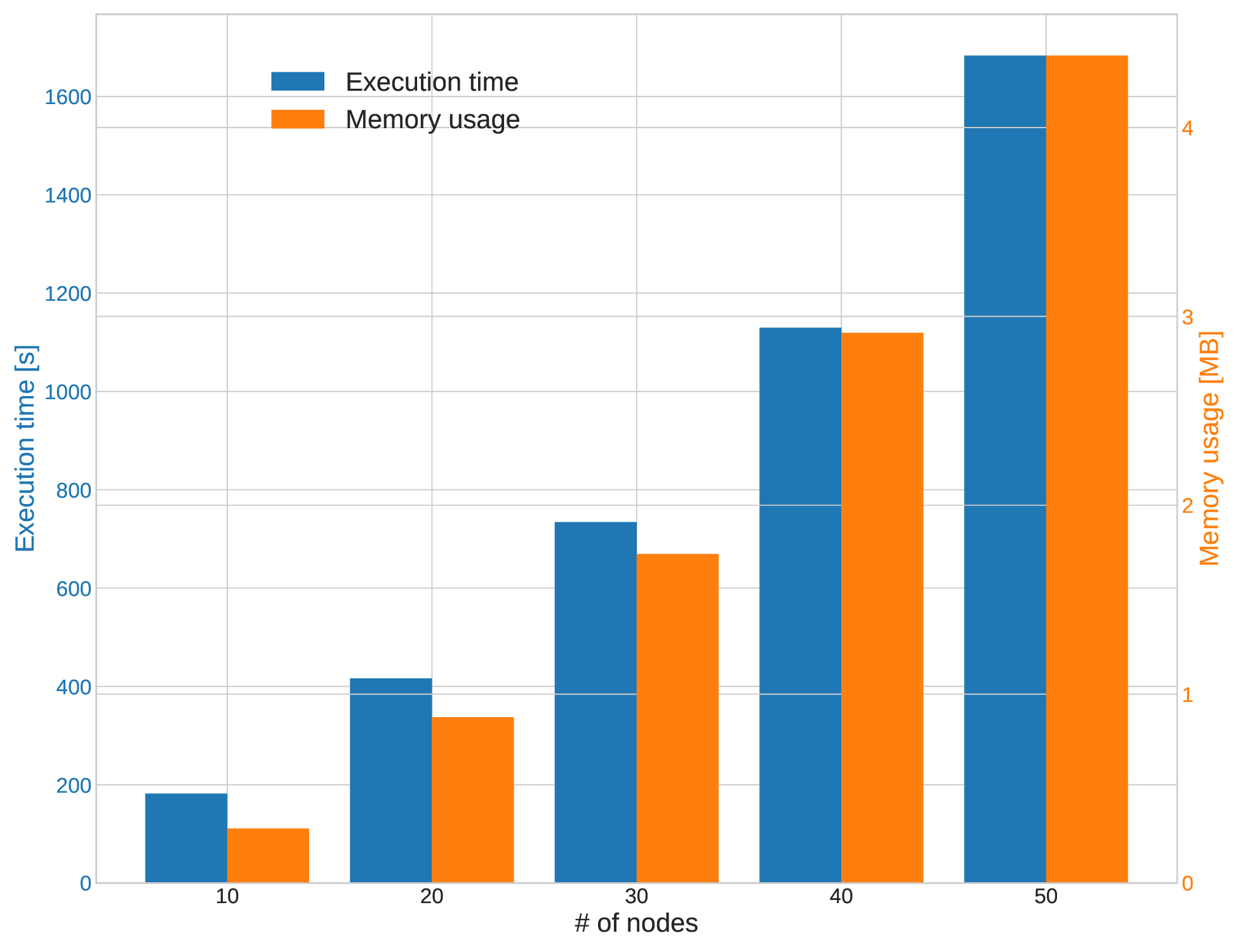}
  \caption{Execution time (primary y-axis, in blue) and memory usage (secondary y-axis, in orange) with varying number of nodes \textit{and} transaction rate.}
  \label{fig:scalability_nodes_txs}
\end{figure}

Finally, Fig.~\ref{fig:scalability_nodes_txs} shows the impact of varying \textit{both} the number of nodes while keeping the same per-node transaction generation rate at one transaction every 12 seconds, which results in increasing \textit{global} transaction rate.
As expected, both execution time and memory usage increase with the combined growth in network size and global transaction rate. In fact, each node must process a larger total number of transactions (leading to higher execution time) and store more transactions data (resulting in increased memory usage). 

\subsection{Comparison against BlockSim Simulator}

In this section, we compare the performance of iblock against BlockSim by Alharby et al.~\cite{alharby2019blocksim}. We consider BlockSim as it offers more functionalities than other existing open-source Bitcoin simulators described in Section~\ref{sec:related_work}.

BlockSim offers two modes of execution, namely \textit{light} and \textit{full}. 
Light mode trades off accuracy for simulation speed, as it does not simulate the different propagation speeds of transactions within the network. 
As a consequence, all simulated nodes view the same mempool, and it is not possible to simulate nodes with different mempools. 
More precisely, the transactions are created just when a new block is mined, so no mempool is actually maintained during the simulation.
Instead, full mode maintains a mempool for each node, so it can simulate the effects of different transaction propagation speed by adding a randomly-generated delay to each transaction before inserting it into the mempool of each node.

Like full-mode BlockSim, iblock considers different propagation times within its simulations, as each transaction is modeled as an OMNeT++ message sent by a source node and received with different random delays by destination nodes.
In addition, iblock can simulate aspects like the variable transaction fees, the UTXOs, and the difficulty adjustments, which are absent in full-mode BlockSim as discussed in Section~\ref{sec:related_work}.
Nevertheless, we chose to compare iblock against the full-mode BlockSim because BlockSim is the most used Bitcoin simulator in the literature, and because its full mode is the most similar to iblock in terms of simulation detail.
In the following, the term `BlockSim' will always refer to its full mode.

We consider a simulation scenario composed of 30 nodes, 10 of which are miners that generate three transactions per second globally. Preliminary simulations on our general-purpose hardware showed that BlockSim failed to complete simulations lasting just a few hours, due to excessive memory usage\footnote{A BlockSim's process simulating a 12-hour scenario was terminated by the operating system after exceeding 30 GB of virtual memory.}. Thus, we limit the simulation duration to six hours (of simulated time).

\begin{figure}[t]
  \centering
  % [trim=left bottom right top, clip]
  \includegraphics[trim=1.5cm 1.5cm 1.5cm 1.5cm, width=\linewidth]{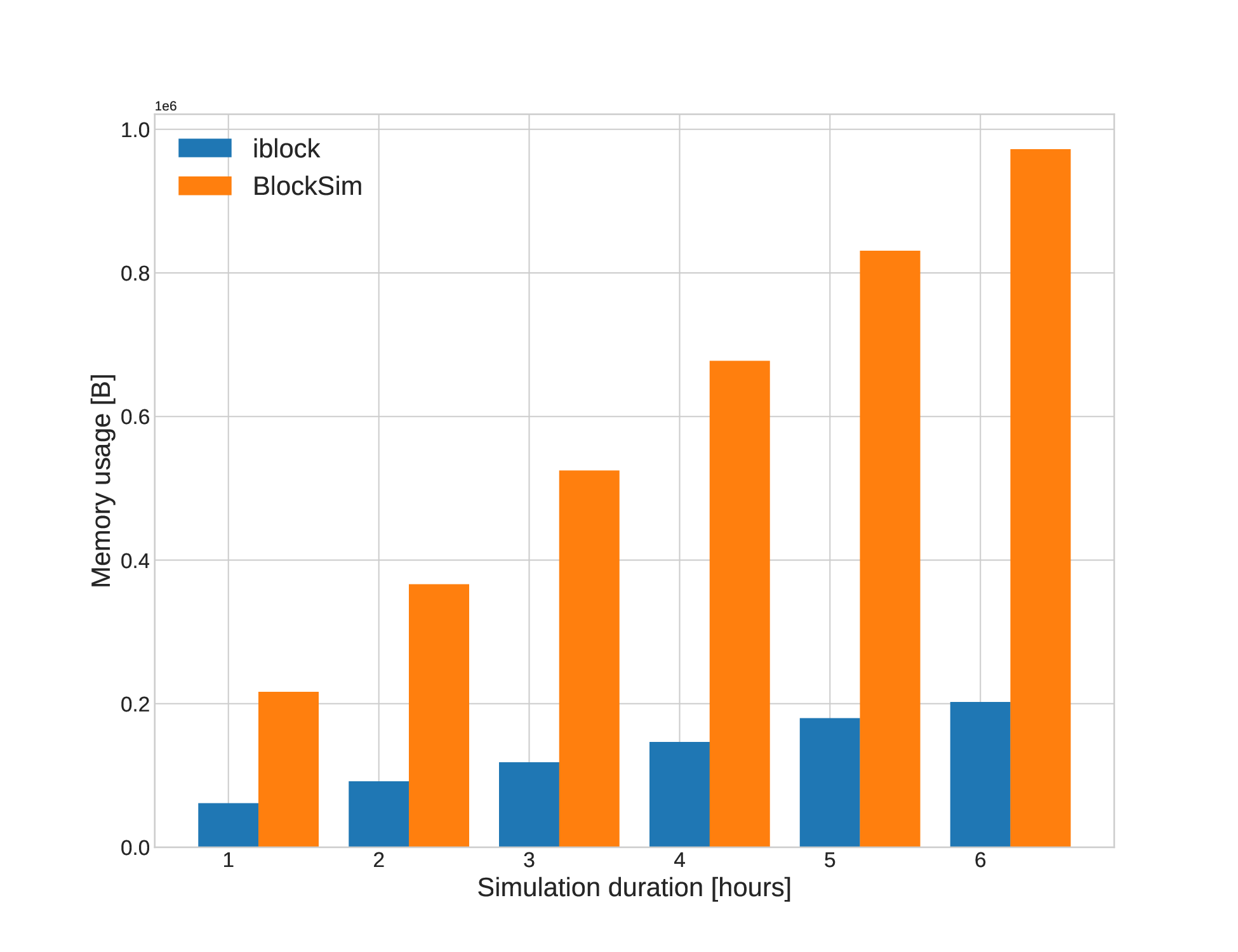}
  \caption{Memory usage comparison between iblock (in blue) and BlockSim (in orange).}
  \label{fig:comparison}
\end{figure}

Fig.~\ref{fig:comparison} shows the total memory consumption for both simulators when increasing the simulation duration from one to six hours, while Table~\ref{tbl:perfcomparison} compares total execution times. Note that the execution time of BlockSim is even higher than the simulated time in the 5-hour and 6-hour scenarios.

The above results demonstrate that iblock achieves significantly better performance than BlockSim in terms of both memory consumption and execution time, while also guaranteeing higher level of simulation detail.

\begin{table}
    \caption{Execution time comparisong between iblock and BlockSim}
    \label{tbl:perfcomparison}
    \centering
    \begin{tabular}{c|c|c}
    Simulated time & iblock execution time & BlockSim execution time \\
     \hline
    1h & 1s & 7m 16s \\
    2h & 2s & 28m 8s  \\
    3h & 4s & 1h 55m 13s \\
    4h & 5s & 3h 13m 56s \\
    5h & 7s & 6h 30m 46s \\
    6h & 10s & 7h 16m 59s \\
\end{tabular}
\end{table}
\section{Conclusions}
\label{sec:conclusions}

This paper proposed iblock, a comprehensive C++ library and simulation model for the Bitcoin system, which enables simulations across various layers, i.e. network, consensus, incentives, data. 
Written in C++ and designed for OMNeT++~\cite{varga2001omnet}, iblock offers superior efficiency and scalability with respect to state-of-the-art simulators written in high-level languages. 
iblock provides in-depth representation of Bitcoin's data structures and algorithms, which facilitates experiments across consensus, incentive mechanisms and data management. 
Moreover, since OMNeT++ is a general-purpose simulator, the possible integration with other OMNeT++ libraries allows highly detailed simulations. 
iblock modular and extensible architecture allows for the seamless addition of new features and behaviors, providing a valuable framework for studying their effects on the Bitcoin system. 
We measured iblock's performance against BlockSim~\cite{alharby2019blocksim}, which is a state-of-the-art blockchain simulator, proving that iblock is more efficient at the same level of simulation detail. 
We also validated iblock's accuracy by comparing simulation results with related studies and real Bitcoin network data, and showcase iblock's capability to simulate different scenarios such as the selfish mine attack~\cite{eyal2013majority}.

\section*{Acknowledgment}
This work was partially supported by the project SERICS (PE00000014) under the MUR National Recovery and Resilience Plan funded by the European Union - NextGenerationEU - AQuSDIT: Advanced and Quantum-safe Solutions for Digital Identity and digital Tracing, CUP H73C22000880001; and by the Italian Ministry of Education and Research (MUR) within the framework of the Forelab project (Departments of Excellence).
%Anonymized acknowledgment.

%\bibliographystyle{unsrtnat}
\bibliographystyle{ACM-Reference-Format}
\bibliography{bibliography} 


\begin{thebibliography}{18}

%%% the Web address.

\ifx \showCODEN    \undefined \def \showCODEN     #1{\unskip}     \fi
\ifx \showDOI      \undefined \def \showDOI       #1{#1}\fi
\ifx \showISBNx    \undefined \def \showISBNx     #1{\unskip}     \fi
\ifx \showISBNxiii \undefined \def \showISBNxiii  #1{\unskip}     \fi
\ifx \showISSN     \undefined \def \showISSN      #1{\unskip}     \fi
\ifx \showLCCN     \undefined \def \showLCCN      #1{\unskip}     \fi
\ifx \shownote     \undefined \def \shownote      #1{#1}          \fi
\ifx \showarticletitle \undefined \def \showarticletitle #1{#1}   \fi
\ifx \showURL      \undefined \def \showURL       {\relax}        \fi
% invisible to TeX
\providecommand\bibfield[2]{#2}
\providecommand\bibinfo[2]{#2}
\providecommand\natexlab[1]{#1}
\providecommand\showeprint[2][]{arXiv:#2}

\bibitem[ine({[n.\,d.]})]%
        {inet}
 \bibinfo{year}{[n.\,d.]}\natexlab{}.
\newblock \bibinfo{booktitle}{\emph{INET Framework}}.
\newblock
\urldef\tempurl%
\url{https://inet.omnetpp.org/}
\showURL{%
\tempurl}


\bibitem[Alharby and van Moorsel(2019)]%
        {alharby2019blocksim}
\bibfield{author}{\bibinfo{person}{Maher Alharby} {and} \bibinfo{person}{Aad
  van Moorsel}.} \bibinfo{year}{2019}\natexlab{}.
\newblock \showarticletitle{{BlockSim}: A Simulation Framework for Blockchain
  Systems}.
\newblock \bibinfo{journal}{\emph{SIGMETRICS Perform. Eval. Rev.}}
  \bibinfo{volume}{46}, \bibinfo{number}{3} (\bibinfo{date}{Jan.}
  \bibinfo{year}{2019}), \bibinfo{pages}{135–138}.
\newblock
\showISSN{0163-5999}
\urldef\tempurl%
\url{https://doi.org/10.1145/3308897.3308956}
\showDOI{\tempurl}


\bibitem[Androulaki et~al\mbox{.}(2013)]%
        {androulaki2013evaluating}
\bibfield{author}{\bibinfo{person}{Elli Androulaki},
  \bibinfo{person}{Ghassan~O. Karame}, \bibinfo{person}{Marc Roeschlin},
  \bibinfo{person}{Tobias Scherer}, {and} \bibinfo{person}{Srdjan Capkun}.}
  \bibinfo{year}{2013}\natexlab{}.
\newblock \showarticletitle{Evaluating User Privacy in {Bitcoin}}. In
  \bibinfo{booktitle}{\emph{Financial Cryptography and Data Security}},
  \bibfield{editor}{\bibinfo{person}{Ahmad-Reza Sadeghi}} (Ed.).
  \bibinfo{publisher}{Springer Berlin Heidelberg}, \bibinfo{address}{Berlin,
  Heidelberg}, \bibinfo{pages}{34--51}.
\newblock
\showISBNx{978-3-642-39884-1}


\bibitem[Antonopoulos and Harding(2023)]%
        {antonopoulos2023mastering}
\bibfield{author}{\bibinfo{person}{Andreas~M. Antonopoulos} {and}
  \bibinfo{person}{David~A. Harding}.} \bibinfo{year}{2023}\natexlab{}.
\newblock \bibinfo{booktitle}{\emph{Mastering Bitcoin: Programming the Open
  Blockchain} (\bibinfo{edition}{3rd} ed.)}.
\newblock \bibinfo{publisher}{O'Reilly Media}, \bibinfo{address}{Sebastopol,
  CA}.
\newblock
\showISBNx{9781098150082}


\bibitem[Aoki et~al\mbox{.}(2019)]%
        {aoki2019simblock}
\bibfield{author}{\bibinfo{person}{Yusuke Aoki}, \bibinfo{person}{Kai Otsuki},
  \bibinfo{person}{Takeshi Kaneko}, \bibinfo{person}{Ryohei Banno}, {and}
  \bibinfo{person}{Kazuyuki Shudo}.} \bibinfo{year}{2019}\natexlab{}.
\newblock \showarticletitle{{SimBlock}: A Blockchain Network Simulator}. In
  \bibinfo{booktitle}{\emph{IEEE INFOCOM 2019 - IEEE Conference on Computer
  Communications Workshops (INFOCOM WKSHPS)}}. \bibinfo{pages}{325--329}.
\newblock
\urldef\tempurl%
\url{https://doi.org/10.1109/INFCOMW.2019.8845253}
\showDOI{\tempurl}


\bibitem[Basile et~al\mbox{.}(2022)]%
        {basile2022segwit}
\bibfield{author}{\bibinfo{person}{M. Basile}, \bibinfo{person}{G. Nardini},
  \bibinfo{person}{P. Perazzo}, {and} \bibinfo{person}{G. Dini}.}
  \bibinfo{year}{2022}\natexlab{}.
\newblock \showarticletitle{{SegWit} Extension and Improvement of the
  {BlockSim} {Bitcoin} Simulator}. In \bibinfo{booktitle}{\emph{2022 IEEE
  International Conference on Blockchain (Blockchain)}}.
  \bibinfo{pages}{115--123}.
\newblock
\urldef\tempurl%
\url{https://doi.org/10.1109/Blockchain55522.2022.00025}
\showDOI{\tempurl}


\bibitem[Cui and Hu(2024)]%
        {cui2024bsela}
\bibfield{author}{\bibinfo{person}{Bo Cui} {and} \bibinfo{person}{Yun Hu}.}
  \bibinfo{year}{2024}\natexlab{}.
\newblock \showarticletitle{{BSELA}: A Blockchain Simulator with Event-Layered
  Architecture}.
\newblock \bibinfo{journal}{\emph{Future Generation Computer Systems}}
  \bibinfo{volume}{151} (\bibinfo{year}{2024}), \bibinfo{pages}{182--195}.
\newblock
\showISSN{0167-739X}
\urldef\tempurl%
\url{https://doi.org/10.1016/j.future.2023.09.034}
\showDOI{\tempurl}


\bibitem[Eyal and Sirer(2018)]%
        {eyal2013majority}
\bibfield{author}{\bibinfo{person}{Ittay Eyal} {and}
  \bibinfo{person}{Emin~G\"{u}n Sirer}.} \bibinfo{year}{2018}\natexlab{}.
\newblock \showarticletitle{Majority is not enough: bitcoin mining is
  vulnerable}.
\newblock \bibinfo{journal}{\emph{Commun. ACM}} \bibinfo{volume}{61},
  \bibinfo{number}{7} (\bibinfo{date}{June} \bibinfo{year}{2018}),
  \bibinfo{pages}{95–102}.
\newblock
\showISSN{0001-0782}
\urldef\tempurl%
\url{https://doi.org/10.1145/3212998}
\showDOI{\tempurl}


\bibitem[Faria and Correia(2019)]%
        {faria2019blocksim}
\bibfield{author}{\bibinfo{person}{Carlos Faria} {and} \bibinfo{person}{Miguel
  Correia}.} \bibinfo{year}{2019}\natexlab{}.
\newblock \showarticletitle{{BlockSim}: Blockchain Simulator}. In
  \bibinfo{booktitle}{\emph{2019 IEEE International Conference on Blockchain
  (Blockchain)}}. \bibinfo{pages}{439--446}.
\newblock
\urldef\tempurl%
\url{https://doi.org/10.1109/Blockchain.2019.00067}
\showDOI{\tempurl}


\bibitem[Gervais et~al\mbox{.}(2016)]%
        {gervais2016security}
\bibfield{author}{\bibinfo{person}{Arthur Gervais}, \bibinfo{person}{Ghassan~O.
  Karame}, \bibinfo{person}{Karl W\"{u}st}, \bibinfo{person}{Vasileios
  Glykantzis}, \bibinfo{person}{Hubert Ritzdorf}, {and} \bibinfo{person}{Srdjan
  Capkun}.} \bibinfo{year}{2016}\natexlab{}.
\newblock \showarticletitle{On the Security and Performance of Proof of Work
  Blockchains}. In \bibinfo{booktitle}{\emph{Proceedings of the 2016 ACM SIGSAC
  Conference on Computer and Communications Security}} (Vienna, Austria)
  \emph{(\bibinfo{series}{CCS '16})}. \bibinfo{publisher}{Association for
  Computing Machinery}, \bibinfo{address}{New York, NY, USA},
  \bibinfo{pages}{3–16}.
\newblock
\showISBNx{9781450341394}
\urldef\tempurl%
\url{https://doi.org/10.1145/2976749.2978341}
\showDOI{\tempurl}


\bibitem[Miller and Jansen(2015)]%
        {miller2015shadowbitcoin}
\bibfield{author}{\bibinfo{person}{Andrew Miller} {and} \bibinfo{person}{Rob
  Jansen}.} \bibinfo{year}{2015}\natexlab{}.
\newblock \showarticletitle{{Shadow-Bitcoin}: Scalable Simulation via Direct
  Execution of Multi-Threaded Applications}. In \bibinfo{booktitle}{\emph{8th
  Workshop on Cyber Security Experimentation and Test (CSET 15)}}.
  \bibinfo{publisher}{USENIX Association}, \bibinfo{address}{Washington, D.C.}
\newblock
\urldef\tempurl%
\url{https://www.usenix.org/conference/cset15/workshop-program/presentation/miller}
\showURL{%
\tempurl}


\bibitem[Nakamoto(2008)]%
        {nakamoto2008bitcoin}
\bibfield{author}{\bibinfo{person}{Satoshi Nakamoto}.}
  \bibinfo{year}{2008}\natexlab{}.
\newblock \showarticletitle{Bitcoin: A Peer-to-Peer Electronic Cash System}.
\newblock  (\bibinfo{year}{2008}).
\newblock
\urldef\tempurl%
\url{https://bitcoin.org/bitcoin.pdf}
\showURL{%
\tempurl}


\bibitem[Nardini et~al\mbox{.}(2020)]%
        {simu5g}
\bibfield{author}{\bibinfo{person}{Giovanni Nardini}, \bibinfo{person}{Dario
  Sabella}, \bibinfo{person}{Giovanni Stea}, \bibinfo{person}{Purvi Thakkar},
  {and} \bibinfo{person}{Antonio Virdis}.} \bibinfo{year}{2020}\natexlab{}.
\newblock \showarticletitle{Simu5G–An OMNeT++ Library for End-to-End
  Performance Evaluation of 5G Networks}.
\newblock \bibinfo{journal}{\emph{IEEE Access}}  \bibinfo{volume}{8}
  (\bibinfo{year}{2020}), \bibinfo{pages}{181176--181191}.
\newblock
\urldef\tempurl%
\url{https://doi.org/10.1109/ACCESS.2020.3028550}
\showDOI{\tempurl}


\bibitem[Paulavičius et~al\mbox{.}(2021)]%
        {paulavicius2021systematic}
\bibfield{author}{\bibinfo{person}{Remigijus Paulavičius},
  \bibinfo{person}{Saulius Grigaitis}, {and} \bibinfo{person}{Ernestas
  Filatovas}.} \bibinfo{year}{2021}\natexlab{}.
\newblock \showarticletitle{A Systematic Review and Empirical Analysis of
  Blockchain Simulators}.
\newblock \bibinfo{journal}{\emph{IEEE Access}}  \bibinfo{volume}{9}
  (\bibinfo{year}{2021}), \bibinfo{pages}{38010--38028}.
\newblock
\urldef\tempurl%
\url{https://doi.org/10.1109/ACCESS.2021.3063324}
\showDOI{\tempurl}


\bibitem[Polge et~al\mbox{.}(2021)]%
        {polge2021blockperf}
\bibfield{author}{\bibinfo{person}{Julien Polge}, \bibinfo{person}{Sankalp
  Ghatpande}, \bibinfo{person}{Sylvain Kubler}, \bibinfo{person}{Jérémy
  Robert}, {and} \bibinfo{person}{Yves Le~Traon}.}
  \bibinfo{year}{2021}\natexlab{}.
\newblock \showarticletitle{{BlockPerf}: A Hybrid Blockchain Emulator/Simulator
  Framework}.
\newblock \bibinfo{journal}{\emph{IEEE Access}}  \bibinfo{volume}{9}
  (\bibinfo{year}{2021}), \bibinfo{pages}{107858--107872}.
\newblock
\urldef\tempurl%
\url{https://doi.org/10.1109/ACCESS.2021.3101044}
\showDOI{\tempurl}


\bibitem[Riebl et~al\mbox{.}(2015)]%
        {artery}
\bibfield{author}{\bibinfo{person}{Raphael Riebl},
  \bibinfo{person}{Hendrik-Jörn Günther}, \bibinfo{person}{Christian Facchi},
  {and} \bibinfo{person}{Lars Wolf}.} \bibinfo{year}{2015}\natexlab{}.
\newblock \showarticletitle{Artery: Extending Veins for VANET applications}. In
  \bibinfo{booktitle}{\emph{2015 International Conference on Models and
  Technologies for Intelligent Transportation Systems (MT-ITS)}}.
  \bibinfo{pages}{450--456}.
\newblock
\urldef\tempurl%
\url{https://doi.org/10.1109/MTITS.2015.7223293}
\showDOI{\tempurl}


\bibitem[Stoykov et~al\mbox{.}(2017)]%
        {stoykov2017vibes}
\bibfield{author}{\bibinfo{person}{Lyubomir Stoykov}, \bibinfo{person}{Kaiwen
  Zhang}, {and} \bibinfo{person}{Hans-Arno Jacobsen}.}
  \bibinfo{year}{2017}\natexlab{}.
\newblock \showarticletitle{{VIBES}: fast blockchain simulations for
  large-scale peer-to-peer networks: demo}. In
  \bibinfo{booktitle}{\emph{Proceedings of the 18th ACM/IFIP/USENIX Middleware
  Conference: Posters and Demos}} (Las Vegas, Nevada)
  \emph{(\bibinfo{series}{Middleware '17})}. \bibinfo{publisher}{Association
  for Computing Machinery}, \bibinfo{address}{New York, NY, USA},
  \bibinfo{pages}{19–20}.
\newblock
\showISBNx{9781450352017}
\urldef\tempurl%
\url{https://doi.org/10.1145/3155016.3155020}
\showDOI{\tempurl}


\bibitem[Varga(2001)]%
        {varga2001omnet}
\bibfield{author}{\bibinfo{person}{Andras Varga}.}
  \bibinfo{year}{2001}\natexlab{}.
\newblock \showarticletitle{The {OMNeT++} discrete event simulation system}. In
  \bibinfo{booktitle}{\emph{Proceedings of the European Simulation
  Multiconference (ESM'2001)}}. SCS European Publishing House,
  \bibinfo{pages}{319--324}.
\newblock


\end{thebibliography}

\end{document}